\documentclass{iopart}
\usepackage[dvips]{graphicx}
\usepackage{amssymb}
\usepackage{color}

\begin{document}

\title{Dilute Bose gas with correlated disorder: A Path Integral Monte Carlo study}
\author{S Pilati$^{1,4}$, S Giorgini$^1$, M Modugno$^2$ and N Prokof'ev$^{3,5}$}

\address{$^1$ Dipartimento di Fisica, Universit\`a di Trento and INO-CNR BEC Center, I-38050 Povo, Trento, Italy}
\address{$^2$ LENS and Dipartimento di Fisica, Universit\`a di Firenze, 50019 Sesto Fiorentino, Italy}
\address{$^3$ Department of Physics, University of Massachusetts, Amherst, MA 01003, USA}
\address{$^4$ Institut f\"ur Theoretische Physik, ETH Zurich, 8093 Zurich, Switzerland}
\address{$^5$ Russian Research Center ``Kurchatov Institute'', 123182 Moscow, Russia}

\ead{pilati@phys.ethz.ch}

\begin{abstract}
We investigate the thermodynamic properties of a dilute Bose gas in a correlated random potential using exact path integral Monte Carlo methods. The study is carried out in continuous space and disorder is produced in the simulations by a 3D speckle pattern with tunable intensity and correlation length. We calculate the shift of the superfluid transition temperature due to disorder and we highlight the role of quantum localization by comparing the critical chemical potential with the classical percolation threshold. 
The equation of state of the gas is determined in the regime of strong disorder, where superfluidity is suppressed and the normal phase exists down to very low temperatures. We find a $T^2$ dependence of the energy in agreement with the expected behavior in the Bose glass phase. We also discuss the major role played by the disorder correlation length and we make contact with a Hartree-Fock mean-field approach that holds valid if the correlation length is very large.  The density profiles are analyzed as a function of temperature and interaction strength. Effects of localization and the depletion of the order parameter are emphasized in the comparison between local condensate and total density. At very low temperature we find that the energy and the particle distribution of the gas are very well described by the $T=0$ Gross-Pitaevskii theory even in the regime of very strong disorder.
\end{abstract}


\maketitle

\section{Introduction}
\label{Introduction}

The dirty boson problem has become a central and fascinating subject in condensed matter physics starting from the first theoretical investigations more than 20 years ago~\cite{Ma, Giamarchi, FWGF}. The interplay between quantum degeneracy, interactions and quenched disorder in a bosonic system gives rise to a rich scenario that exhibits new and peculiar features compared to the much older problem of the metal/insulator transition with electrons~\cite{Mott, Anderson, Belitz}. An important difference is that bosons can not rely on the Pauli pressure and repulsive interactions are crucial to avoid collapse in the lowest localized single-particle state. As a result perturbation schemes starting from the non-interacting model, that are most useful for fermions, are completely inappropriate in the case of bosons.  

Theoretical investigations, including quantum Monte Carlo simulations, have mainly addressed the problem of bosons on a lattice with on site bound disorder, the so-called disordered Bose-Hubbard model. In this case commensurability, {\it i.e.} the integer ratio of the number of particles to the number of lattice sites, plays a major role allowing for a superfluid/insulator (of the Mott type) transition also in the absence of disorder that is purely driven by interaction effects. Furthermore, depending on the value of the interaction strength, disorder can act in favor of superfluidity, by randomizing the insulating state close to the Mott transition, or in opposition to it by localizing almost free particles into single-particle levels. The insulating phases occurring in the two regimes of strong and weak interactions are respectively often referred to as the Bose glass, when interactions suppress superfluidity, and the Anderson glass, when interactions compete with disorder enhancing superfluidity~\cite{Scalettar, Gurarie}.  The disorder driven quantum phase transition occurring at $T=0$ has been investigated in a series of numerical studies both at incommensurate and commensurate densities and in various dimensionalities~\cite{Scalettar, Krauth, Sorensen, Makivic, Zhang, Prokofev, Hitchcock, Gurarie}. The picture emerging from these studies, together with the crucial role of interactions to stabilize the system, is that superfluidity is lost for strong enough disorder, leading to a gapless normal phase different from the incompressible Mott insulator. 

Random potentials in continuous systems have been considered using perturbative approaches based on the Bogoliubov theory~\cite{Huang, GPS, Kobayashi, Lopatin, Falco}. These methods are reliable when both interactions and disorder are weak and allow for the determination of the effect of disorder on the thermodynamic properties, including the fraction of atoms in the condensate, the superfluid density and other thermodynamic functions. Exact numerical methods have also been applied both at zero~\cite{Astra} and at finite temperature~\cite{Boninsegni, Gordillo}. In particular, the path-integral Monte Carlo simulations carried out at finite $T$ addressed the problem of the elementary excitations~\cite{Boninsegni} and of the transition temperature~\cite{Gordillo} of a Bose fluid in a random environment. In the case of the continuous-space liquid phase, disorder always acts against superfluidity, whereas interaction helps make the superfluid state more robust. For strong disorder and low temperatures one expects the system to enter an insulating phase (Bose glass) that is smoothly connected with the high-temperature normal phase existing when the disorder is weak.     

On the experimental side a large body of work was devoted to $^4$He adsorbed in porous media, such as Vycor glass and aerogels. These studies investigated the behavior of the heat capacity and of the superfluid response~\cite{Reppy1, Reppy2, Reppy3}, as well as the dynamic structure factor~\cite{Glyde1, Glyde2} as a function of temperature and filling. A suppression of the $\lambda$ transition is observed and the critical coverage for the onset of superfluidity is determined as a function of temperature, however, no clear evidence is found of a compressible Bose glass phase. More recently the dirty boson problem has been addressed using ultracold atoms, which offer unprecedented control and tunability of the disorder parameters and of the interaction strength. Transport and phase-coherence properties of an interacting gas in disordered optical potentials are investigated and an insulating state is reached by increasing the strength of disorder~\cite{Florence1, Fort, Paris1, Florence2, Hulet, DeMarco1, DeMarco2}. A large experimental effort has also been devoted to the suppression of diffusion for non-interacting particles (Anderson localization)~\cite{Florence3, Paris2}.

In this work we report on a path-integral Monte Carlo (PIMC) study of an interacting Bose gas in the presence of correlated disorder produced by 3D optical speckles. This random potential is relevant for experiments and allows for an independent tuning of intensity and correlation length. By increasing the disorder strength, we find a sizable reduction of the superfluid transition temperature and the shift is larger for weaker interactions. We map out the normal to superfluid phase diagram, both in the chemical potential vs. disorder and in the density vs. disorder plane. For strong disorder and in the presence of small but finite interactions, the critical chemical potential varies linearly with the disorder intensity and is essentially independent of temperature and interaction strength, in agreement with the existence of a mobility edge separating localized from extended states. We also find that the critical chemical potential is much larger than the classical percolation threshold for the same random potential, implying that a major role is played by quantum localization effects. In the regime of strong disorder and for chemical potentials below the critical value, the equilibrium state is a highly degenerate normal gas. We investigate the thermodynamic properties of this phase, finding a $T^2$ dependence of the equation of state in agreement with the peculiar feature expected for the Bose glass phase. The effect of the disorder correlation length is discussed in detail and we show that a non-trivial behavior is obtained only when the correlation length is comparable with the mean interparticle distance. At $T=0$ we also carry out calculations using the Gross-Pitaevskii (GP) equation and at finite $T$ using a self-consistent mean-field approach based on Hartree-Fock theory and on the local density approximation. The results of the GP equation for the ground-state energy and the spatial distribution of particles are accurate even in the regime of strong disorder with short-range correlations. This conclusion might be useful in view of investigating the structural properties of the Bose glass phase.    

We consider a system of $N$ identical bosons of mass $m$, subject to the random field $V_{\rm{dis}}$ and interacting with a repulsive pairwise potential. The Hamiltonian of the gas takes then the form:
\begin{equation}
\hat{H}=\sum_{i=1}^N \left(-\frac{\hbar^2}{2m}\nabla_i^2+V_{\rm{dis}}({\bf r}_i)\right)+
\sum_{i<j}V(|{\bf r}_i-{\bf r}_j|) \;.
\label{Intro1}
\end{equation}
Interatomic interactions are modeled using a hard-sphere potential: 
\begin{equation}
V(r)=\left\{  \begin{array}{cc} +\infty  & (r<a)  \\
                                      0  & \;\; (r>a)\;,  \end{array} \right.
\label{Intro2}
\end{equation}
where the diameter $a$ coincides with the $s$-wave length of the two-body scattering problem. Furthermore, the system is in a cubic box of volume $\Omega=L^3$ with periodic boundary conditions.

The structure of the paper is as follows. In section~\ref{Section1} we introduce the random potential and its statistical properties. In section~\ref{Section2} we discuss classical percolation for the speckle potential and we estimate the percolation threshold in 3D. Some details of the PIMC numerical method are presented in section~\ref{Section3}. In section~\ref{Section4} we report our results on the superfluid transition: shift of the critical temperature, critical chemical potential and critical density. Most of these results were already presented in a previous publication of some of us~\cite{PRL}. In section~\ref{Section5} we introduce a mean-field approach based on the GP equation at $T=0$ and on a Hartree-Fock self-consistent theory at finite $T$ and for long-correlated disorder. The low temperature thermodynamics is discussed in section~\ref{Section6}, including the equation of state, the condensate and total density profiles and the behavior of superfluid density and condensate fraction as a function of temperature and interaction strength in the disordered phase. Finally, we draw our conclusions in section~\ref{Conclusions}.

\begin{figure}
\begin{center}
\includegraphics[width=8cm]{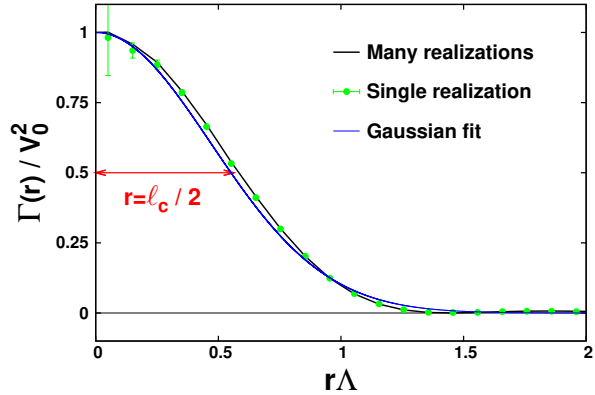}
\caption{(color online). Radial dependence (in units of the inverse momentum cut-off $\Lambda$) of the disorder spatial autocorrelation function $\Gamma$. The solid (black) line refers to an average over many realizations of the random field, the (green) symbols correspond to a single realization. The (blue) line is a Gaussian fit.}
\label{fig1}
\end{center}
\end{figure}

\begin{figure}
\begin{center}
\includegraphics[width=8cm]{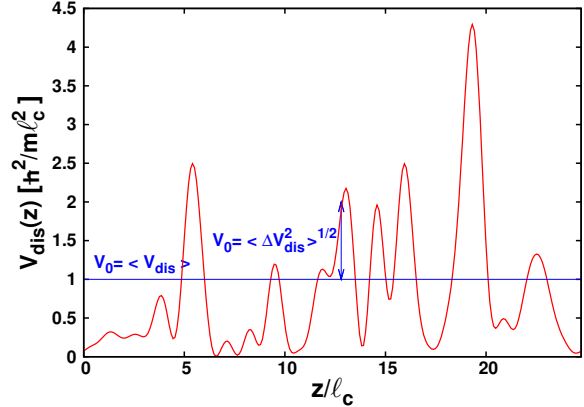}
\caption{(color online). Typical shape of the speckle potential $V_{\rm{dis}}$, with averaged value $V_0=\hbar^2/m\ell_c^2$, shown in the direction (0,0,1) of the simulation box.}
\label{fig2}
\end{center}
\end{figure}

\begin{figure}
\begin{center}
\includegraphics[width=8cm]{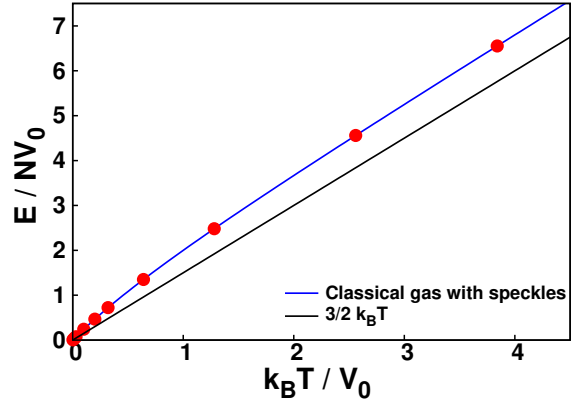}
\caption{(color online). Energy per particle of a classical non-interacting gas subject to the speckle potential. The homogeneous value $E/N=3k_BT/2$ is also shown as a reference.}
\label{fig3}
\end{center}
\end{figure}

\section{Speckle potential}
\label{Section1}

The random external field we consider is the one produced by 3D optical speckles. The local intensity is obtained from the following expression~\cite{Huntley}:
\begin{equation}
V_{\rm{dis}}({\bf r})=V_0\biggl|\frac{1}{\Omega} \int d{\bf k} \tilde{\varphi}({\bf k})W({\bf k}) e^{i{\bf k}\cdot{\bf r}}\biggr|^2 \;,
\label{speckle1}
\end{equation}
where $V_0$ is a positive constant and
\begin{equation}
\tilde{\varphi}({\bf k})=\int d{\bf r} \varphi({\bf r}) e^{-i{\bf k}\cdot{\bf r}}
\label{speckle2}
\end{equation}
is the Fourier transform of the complex field $\varphi({\bf r})$, whose real and imaginary part are independent random variables sampled from a gaussian distribution with zero mean and unit variance. The function $W({\bf k})$ is a low-wavevector filter defined as
\begin{equation}
W({\bf k})=\left\{  \begin{array}{cc} 1  & (k<\pi\Lambda)  \\
                                      0  & \;\; (k>\pi\Lambda)\;.  \end{array} \right.
\label{speckle3}
\end{equation}
The random potential in equation~(\ref{speckle1}) is positive definite and the probability distribution of intensities is given by the normalized exponential law~\cite{Goodman} 
\begin{equation}
P(V_{\rm{dis}})=\frac{1}{V_0}e^{-V_{\rm{dis}}/V_0} \;. 
\label{speckle4}
\end{equation}
If the system's size is large enough the above defined disordered potential is self-averaging, {\it i.e.} spatial averages coincide with averages over different realizations. For a generic function $f(V_{\rm{dis}})$ of the disorder intensity one can thus write the following identity 
\begin{equation}
\frac{1}{\Omega}\int d{\bf r} f[V_{\rm{dis}}({\bf r})]=\int_0^\infty dV_{\rm{dis}} P(V_{\rm{dis}}) 
f(V_{\rm{dis}}) \equiv \langle f(V_{\rm{dis}})\rangle \;. 
\label{speckle5}
\end{equation}
According to this property, the spatial average and the corresponding root-mean-square displacement of the speckle potential are both determined by the energy scale $V_0$: $\langle V_{\rm{dis}}\rangle=V_0$ and $\sqrt{\langle V_{\rm{dis}}^2\rangle-\langle V_{\rm{dis}}\rangle^2}=V_0$. The correlation length $\ell_c$ of the random field is defined from the spatial autocorrelation function, 
\begin{equation}
\Gamma(r)=\langle V_{\rm{dis}}({\bf r}^\prime)V_{\rm{dis}}({\bf r}^\prime+{\bf r})\rangle-\langle V_{\rm{dis}}\rangle^2
\label{speckle6}
\end{equation}
as the length scale for which $\Gamma(\ell_c/2)=\Gamma(0)/2$. 

The above equations characterizing the speckle intensity field in 3D can be straightforwardly generalized to 2D and 1D. In particular, in 1D the autocorrelation function $\Gamma(x)$ takes the simple form~\cite{Goodman}
\begin{equation}
\Gamma(x)=\left(\frac{\sin(\pi\Lambda x)}{\pi\Lambda x}\right)^2 \;,
\label{speckle7}
\end{equation}
and $\ell_c=0.88/\Lambda$. In 3D we find that the autocorrelation function $\Gamma(r)$ is well approximated by a gaussian (see figure~\ref{fig1}) and we obtain numerically $\ell_c=1.1/\Lambda$. It is important to notice that standard experimental realizations of optical speckles are 2D, {\it i.e.} the speckle pattern lies in the plane perpendicular to the propagation of the laser beam, featuring equal correlation lengths in the $x$ and $y$ planar directions and a much larger $\ell_c$ in the $z$ direction. We consider instead a 3D pattern, having the same correlation length in the three spatial directions. This random field can be realized, for example, by adding speckle patterns generated from different directions. 

The typical shape of the speckle potential $V_{\rm{dis}}$ is shown in figure~\ref{fig2}: typical wells have size $\ell_c$ and depth $V_0$. The energy $\hbar^2/m\ell_c^2$, associated with the correlation length $\ell_c$, and $V_0$ provide the two relevant energy scales for the disorder potential. In particular, if $V_0\gg\hbar^2/m\ell_c^2$ the random potential is classical in nature, with typical wells that are deep enough to sustain many single-particle bound states. The opposite regime, $V_0\ll\hbar^2/m\ell_c^2$, corresponds instead to quantum disorder, where typical wells of size $\ell_c$ do not have bound states and these can be supported only by rare wells of size much larger than $\ell_c$ or with depth much larger than $V_0$.

The root-mean-square intensity $V_0$ and the correlation length $\ell_c$ are the relevant parameters characterizing in general the various types of disorder. For example, the delta-correlated disorder, which has been considered in many theoretical investigations~\cite{Huang,Lopatin,Falco,Yukalov}, is defined by the following autocorrelation function:
\begin{equation}
\langle\Delta V_{\rm{dis}}({\bf r})\Delta V_{\rm{dis}}({\bf r}^\prime)\rangle=\kappa\delta({\bf r}-{\bf r}^\prime) \;,
\label{speckle8}
\end{equation}
where $\Delta V_{\rm{dis}}({\bf r})=V_{\rm{dis}}({\bf r})-\langle V_{\rm{dis}}\rangle$ is the displacement from the average. By approximating the speckle $\Gamma$ function~(\ref{speckle6}) using a gaussian function, $\Gamma(r)=V_0^2e^{-r^2/2\sigma^2}$, with $\sigma=\ell_c/\sqrt{8\log2}$ to recover the same half width at half maximum, one finds that the speckle field in the limit $\ell_c\to0$ reproduces a delta-correlated disorder with the strength $\kappa$ given by
\begin{equation}
\kappa=\left(\frac{\pi}{4\log2}\right)^{3/2}V_0^2\ell_c^3 \;.
\label{speckle9}
\end{equation}
  
In our simulations the length scale $\ell_c$ is typically $\sim 100$ times larger than the hard-sphere diameter $a$, allowing for a wide range of disorder intensities where interaction effects are well described by the $s$-wave scattering length and the details of the interatomic potential are irrelevant. The typical box size used in the simulations ranges from $L\sim20\ell_c$ to $L\sim50\ell_c$. An indication of self-averaging of disorder for these values of $L$ is provided by figure~\ref{fig1}, where we show the comparison between the autocorrelation function $\Gamma$ averaged over many realizations of the random potential and the one corresponding to a single realization.  

The self-averaging property (\ref{speckle5}) allows one to calculate the thermodynamics of a classical non-interacting gas. For example, the average energy per particle obtained from the spatial average of the disordered potential over the Boltzmann factor 
\begin{equation}
\frac{E}{N}=\frac{3}{2}k_BT+\frac{\int d{\bf r} V_{\rm{dis}}({\bf r}) e^{-V_{\rm{dis}}({\bf r})/k_BT}}
{\int d{\bf r} e^{-V_{\rm{dis}}({\bf r})/k_BT}} \;,
\label{speckle10}
\end{equation}
yields the following simple result 
\begin{equation}
\frac{E}{N}=\frac{3}{2}k_BT+\frac{V_0}{1+V_0/k_BT} \;.
\label{speckle11}
\end{equation}
In figure~\ref{fig3} we compare the above analytical prediction with the results obtained from a direct spatial integration using a typical size of the simulation box. The good agreement found shows that in our simulations the self-averaging property is well satisfied for non-trivial functions of the disorder intensity.

\begin{figure}
\begin{center}
\includegraphics[width=8cm]{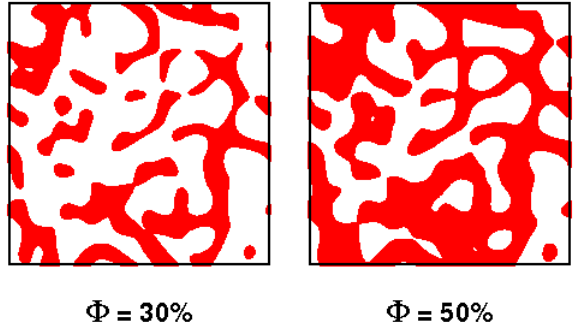}
\caption{(color online). Percolation in a 2D speckle potential: the left and right panel correspond respectively to an accessible volume (shown in red) below and above the percolation threshold.}
\label{fig4}
\end{center}
\end{figure}

\begin{figure}
\begin{center}
\includegraphics[width=8cm]{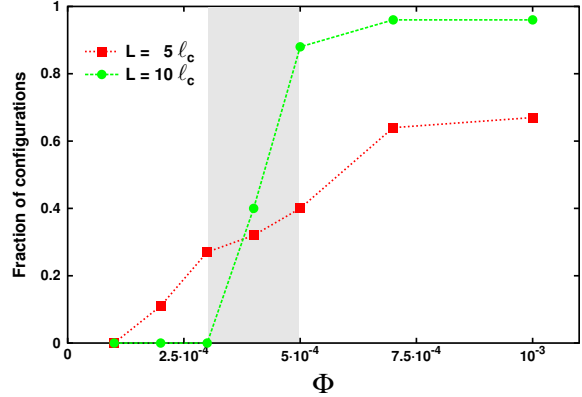}
\caption{(color online). Fraction of configurations of a 3D speckle pattern where percolation occurs in one of the spatial directions as a function of the accessible volume. Results are shown for two different system sizes. The estimated percolation threshold is $4(1)\cdot10^{-4}$ and is shown with the shaded area. }
\label{fig5}
\end{center}
\end{figure}

\section{Classical percolation}
\label{Section2}

In this Section we investigate the problem of the conducting/insulating transition in a speckle potential from the point of view of classical percolation. The relevant question is to determine the mobility edge of a classical particle subject to the random field~\cite{Zallen}. 

Given the disordered potential $V_{\rm{dis}}({\bf r})$, the fraction of space accessible to particles of energy $\epsilon$ is defined as the fractional volume where the reference energy exceeds the external field:
\begin{equation}
\Phi(\epsilon)=\frac{1}{\Omega}\int_{V_{\rm{dis}}({\bf r})<\epsilon}d{\bf r} \;.
\label{percol1}
\end{equation}
The percolation threshold corresponds to the critical value $\Phi_c$ of the fractional volume such that, if $\Phi(\epsilon)>\Phi_c$, there are infinitely extended volumes of allowed space and particles having energy $\epsilon$ can move across the whole system. In terms of energy, the value $\epsilon_c$ determines the percolation threshold: $\Phi(\epsilon_c)=\Phi_c$. It corresponds to the classical mobility edge separating localized states with energy $\epsilon<\epsilon_c$ from delocalized ones with energy $\epsilon>\epsilon_c$. In the case of speckles the function $\Phi(\epsilon)$ can be simply expressed in terms of the disorder intensity $V_0$ using the property (\ref{speckle5}). One finds
\begin{equation}
\Phi(\epsilon)=1-e^{-\epsilon/V_0} \;.
\label{percol2}
\end{equation}

The determination of the percolation threshold for lattice or continuum models is in general a difficult numerical task and the study of percolation has become a mature branch of statistical physics (see {\it e.g.} \cite{Stauffer}). Our discussion here is limited to the calculation of $\Phi_c$ and of the corresponding thereshold energy $\epsilon_c$ for the speckle potential. The role of dimensionality is crucial for this problem. In 1D the unbound nature of the disordered field implies that, approaching the thermodynamic limit, potential barriers occur higher than any finite energy $\epsilon$ and consequently $\epsilon_c=+\infty$ and $\Phi_c=1$. In 2D the percolation threshold of laser speckles was investigated experimentally~\cite{Smith} using photolithography on a conducting film obtaining the value $\Phi_c=0.407$. This result was later confirmed by a numerical study~\cite{Weinrib}. To our knowledge there are no determination of $\Phi_c$ for 3D speckle patterns.

We estimate the percolation threshold in the continuum by mapping the accessible and unaccessible regions of space on a finite grid. This is done by simply comparing the local value of the external field $V_{\rm{dis}}({\bf r}_i)$ at the grid point ${\bf r}_i$ with the reference energy $\epsilon$. One then investigates the percolation of accessible grid points of the corresponding matrix. In figure~\ref{fig4} we show two typical configurations in the case of 2D speckles: panel a) corresponds to $\Phi=0.30$ well below the percolation threshold and panel b) to $\Phi=0.50$ where percolation, {\it i.e.} existence of an uninterrupted path of accessible points across the whole system in at least one of the spatial directions, has clearly occurred. Two issues have to be considered with care: i) the size $L$ in units of the correlation length $\ell_c$ must be increased in order to extrapolate to the thermodynamic limit and ii) for a given size $L$ the grid points must be dense enough. In figure~\ref{fig5} we plot the fraction of configurations exhibiting percolation in 3D speckle patterns as a function of the accessible volume $\Phi$ and for different system sizes. An estimate of the threshold gives $\Phi_c\simeq\epsilon_c/V_0=4(1)\cdot10^{-4}$. For comparison, we estimated the percolation threshold in 2D obtaining the value $\Phi_c\simeq0.4$ in agreement with previous determinations~\cite{Smith,Weinrib}. The value found for $\Phi_c$ in 3D is rather small, suggesting the presence of many long valleys where particles with a tiny fraction of the energy $V_0$ can freely move across the whole system. One should notice that a similar 3D continuum model, the percolation of voids between overlapping spheres (the so called ``Swiss cheese'' model), gives the much larger value $\Phi_c\simeq0.03$~\cite{Kertesz,Elam} for the percolation threshold.

\section{PIMC method}
\label{Section3}
The partition function $Z$ of a bosonic system with inverse temperature $\beta=(k_BT)^{-1}$ is defined as the trace over all states of the density matrix $\hat{\rho}=e^{-\beta \hat{H}}$ properly symmetrized. 
The partition function satisfies the  convolution equation
\begin{eqnarray}
Z &=& \frac{1}{N!}\sum_P \int d{\bf R} \rho({\bf R},P{\bf R},\beta) = 
\frac{1}{N!}\sum_P \int d{\bf R} 
\label{PIMC1}\\ \nonumber
&\times&  \int d{\bf R}_2 ... \int d{\bf R}_M \rho({\bf R},{\bf R}_2,\tau)...
\rho({\bf R}_M,P{\bf R},\tau) \;,
\end{eqnarray}
where $\tau=\beta/M$, ${\bf R}$ collectively denotes the position vectors ${\bf R}=({\bf r}_1,{\bf r}_2,...,{\bf r}_N)$, $P{\bf R}$ denotes the position vectors with permuted labels  $P{\bf R}=({\bf r}_{P(1)},{\bf
r}_{P(2)},...,{\bf r}_{P(N)})$ and the sum extends over the $N!$ permutations of $N$  particles. The calculation of the partition function in equation~(\ref{PIMC1}) can be mapped to a classical-like simulation of polymeric chains with a number of beads $M$ equal to the number of terms of the convolution 
integral. In a PIMC calculation, one makes use of suitable approximations for the density matrix  $\rho({\bf R},{\bf R}^\prime,\tau)$ at the higher temperature $1/\tau$ in equation~(\ref{PIMC1}) and performs the  multidimensional integration over ${\bf R}$, ${\bf R}_2$,...,${\bf R}_M$ as well as the sum over permutations $P$ by  Monte Carlo sampling~\cite{Ceperley}. The statistical expectation value of a
given operator $O({\bf R})$, 
\begin{equation}
\langle O\rangle = \frac{1}{Z}\frac{1}{N!}\sum_P 
\int d{\bf R} O({\bf R}) \rho({\bf R},P{\bf R},\beta) \;,
\label{PIMC2}
\end{equation}     
is calculated by generating stochastically a set of configurations $\{{\bf R}_i\}$, sampled from a probability density  proportional to the symmetrized density matrix, and then by averaging over the set of values
$\{O({\bf R}_i)\}$. 

An approximation for the high temperature density matrix, which is particularly well suited for studies of dilute gases, is based on the pair-product ansatz~\cite{Ceperley}
\begin{equation}
\rho({\bf R},{\bf R}^\prime,\tau)=\prod_{i=1}^N\rho_1({\bf r}_i,{\bf r}_
i^\prime,\tau)\prod_{i<j}
\frac{\rho_{rel}({\bf r}_{ij},{\bf r}_{ij}^\prime,\tau)}
{\rho_{rel}^0({\bf r}_{ij},{\bf r}_{ij}^\prime,\tau)} \;.
\label{PIMC3}
\end{equation}  
In the above equation $\rho_1$ is the single-particle ideal-gas density matrix
\begin{equation}
\rho_1({\bf r}_i,{\bf r}_i^\prime,\tau)=\left(\frac{m}{2\pi\hbar^2\tau}
\right)^{3/2} 
e^{-({\bf r}_i-{\bf r}_i^\prime)^2m/(2\hbar^2\tau)} \;,
\label{PIMC4}
\end{equation}
and $\rho_{rel}$ is the two-body density matrix of the interacting system, which depends on the relative coordinates  ${\bf r}_{ij}={\bf r}_i-{\bf r}_j$ and ${\bf r}_{ij}^\prime={\bf r}_i^\prime-{\bf r}_j^\prime$,  
divided by the corresponding ideal-gas term
\begin{equation}
\rho_{rel}^0({\bf r}_{ij},{\bf r}_{ij}^\prime,\tau)=
\left( \frac{m}{4\pi\hbar^2\tau} \right)^{3/2} 
e^{-({\bf r}_{ij}-{\bf r}_{ij}^\prime)^2 m/(4\hbar^2\tau)} \;.
\label{PIMC5}
\end{equation}

The two-body density matrix at the inverse temperature $\tau$, $\rho_{rel}({\bf r},{\bf r}^\prime,\tau)$, can be calculated  for a given spherical potential $V(r)$ using the partial-wave decomposition 
\begin{eqnarray}
\rho_{rel}({\bf r},{\bf r}^\prime,\tau)&=&\frac{1}{4\pi}
\sum_{\ell=0}^\infty (2\ell +1)P_\ell(\cos\theta) 
\label{PIMC6} \\ \nonumber
&\times&\int_0^\infty dk e^{-\tau\hbar^2k^2/m} R_{k,\ell}(r)R_{k,\ell}(r^\prime) \;,
\end{eqnarray} 
where $P_\ell(x)$ is the Legendre polynomial of order $\ell$ and $\theta$ is the angle between ${\bf r}$ and ${\bf r}^\prime$. The functions $R_{k,\ell}(r)$ are solutions of the radial Schr\"odinger equation 
\begin{eqnarray}
&-&\frac{\hbar^2}{m}\left( \frac{d^2R_{k,\ell}}{dr^2} +\frac{2}{r} \frac{dR_{k,\ell}}{dr} 
-\frac{\ell(\ell+1)}{r^2}R_{k,\ell}\right) \nonumber\\
&+& V(r)R_{k,\ell} = \frac{\hbar^2k^2}{m}R_{k,\ell} \;,
\label{PIMC7}
\end{eqnarray}
with the asymptotic behavior
\begin{equation}
R_{k,\ell}(r)=\sqrt{\frac{2}{\pi}}\frac{\sin(kr-\ell\pi/2+\delta_\ell)}{r} \;,
\label{PIMC8}
\end{equation}
holding for distances $r\gg R_0$, where $R_0$ is the range of the potential. The phase shift $\delta_\ell$ of the  partial wave of order $\ell$ is determined from the solution of equation~(\ref{PIMC7}) for the given
interatomic potential $V(r)$. 

For the hard-sphere potential (\ref{Intro2}) a simple analytical approximation of the high-temperature two-body density matrix due to Cao and  Berne~\cite{Cao} has been proven to be highly accurate~\cite{PSBCG}. The Cao-Berne density matrix $\rho_{rel}^{CB}({\bf r},{\bf r}^\prime,\tau)$ is obtained using the large momentum expansion of the hard-sphere phase shift  $\delta_\ell\simeq-ka+\ell\pi/2$ and the large momentum expansion of the solutions of the Sch\"odinger equation~(\ref{PIMC7}) $R_{k,\ell}(r)\simeq\sqrt{2/\pi}\sin[k(r-a)]/r$. This yields the result
\begin{eqnarray}
\frac{\rho_{rel}^{CB}({\bf r},{\bf r}^\prime,\tau)}
{\rho_{rel}^0({\bf r},{\bf r}^\prime,\tau)}&=& 
1 -\frac{a(r+r^\prime)-a^2}{rr^\prime} \\ \nonumber
&\times& e^{-[rr^\prime +a^2-a(r+r^\prime)](1+\cos\theta)m/(2\hbar^2\tau)} \;.
\label{PIMC9}
\end{eqnarray}

Simulations are based on the worm algorithm~\cite{BPS}, which allows for an efficient sampling of permutation cycles. In this scheme one samples both diagonal configurations, contributing to averages of the type (\ref{PIMC2}) where all paths are closed,  and off-diagonal configurations where one path is open. These latter configurations contribute to the one-body density matrix (OBDM) defined as 
\begin{equation}
n_1({\bf r}_1,{\bf r}_1^\prime)
= \frac{N}{Z}\frac{1}{N!}\sum_P \int d{\bf r}_2\cdot\cdot\cdot d{\bf r}_N \rho({\bf R},P{\bf R},\beta) \;, 
\label{PIMC10}
\end{equation}
where ${\bf r}_{P(1)}={\bf r}_1^\prime$. The long-range behavior of the OBDM determines the condensate density
\begin{equation}
n_0=\lim_{|{\bf r}-{\bf r}^\prime|\to\infty}n_1({\bf r},{\bf r}^\prime) \;.
\label{PIMC11}
\end{equation}
In our simulations the largest displacement of the OBDM we consider is $|{\bf r}-{\bf r}^\prime|=L/2$. If the size $L$ is large enough the number $N_0$ of condensate particles can be written as 
\begin{equation}
N_0=\int d{\bf r} n_1({\bf r},{\bf r}^\prime) \;, 
\label{PIMC12}
\end{equation} 
where ${\bf r}^\prime$ is fixed by the constraint $|{\bf r}-{\bf r}^\prime|=L/2$ and we perform an average over the solid angle. The quantity under the integral corresponds to the local condensate density at position ${\bf r}$, which could be highly non uniform in the presence of a random potential. 

Beside the condensate density $n_0$, in the present study we consider also the superfluid density $\rho_s$. The superfluid component is the part of the fluid that remains at rest when an infinitely slow movement is applied to the walls that contain the system. In the path-integral formalism, the superfluid fraction of a fluid contained in a box with periodic boundary conditions can be related\cite{Ceperley} to the fluctuations of the \emph{winding number} via the equation
\begin{equation}
\label{rhos}
\frac{\rho_s}{\rho} = \frac{m\langle{\bf W}^2 \rangle}{3\hbar^2 \beta N}.
\end{equation}
The \emph{winding number} ${\bf W}$ is defined as:
\begin{equation}
\label{windingnumber}
{\bf W} = \sum_{i=1}^N  \sum_{m=1}^M \left( {\bf r}_{m+1}^i - {\bf r}_m^i \right).
\end{equation}
It is a topological property of the configuration. It counts the net number of paths that cross any plane perpendicular to one axis. The worm algorithm is particularly suitable to perform ergodic random walks that span all possible winding numbers since it extends the configurations space by including configurations with an open path. Only the Monte Carlo moves that modify the open path can efficiently change the winding number.

We perform calculations both in the canonical (at fixed density $n$) and in the grand-canonical ensemble (at fixed chemical potential $\mu$)~\cite{BPS}. We supplement the worm algorithm with two additional Monte Carlo updates that change the particle number $N$. The first update adds one particle to the system by placing a closed path at a randomly selected position. The second update erases a randomly selected  closed path. The acceptance probability of the first (second) update is fixed by the fugacity $e^{\beta \mu}$ (by its inverse), by the change in the interaction energy due to the path to be inserted (erased) and by the factor $\frac{\Omega C}{N+1}$ ($\frac{N}{\Omega C}$) that takes into account the density change and the normalization of the free particle propagator $C\equiv\left(2\pi\hbar^2\beta/m\right)^{-\frac{3}{2}}$.

\begin{figure}
\begin{center}
\includegraphics[width=8cm]{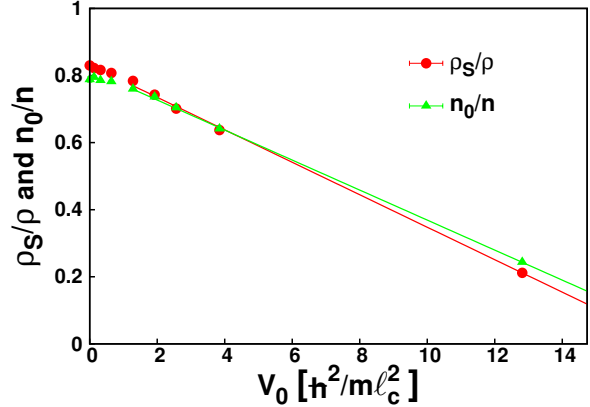}
\caption{(color online). Superfluid density and condensate fraction as a function of the disorder intensity $V_0$. The particle density is here $na^3=10^{-4}$ and the temperature is $T=0.5T_c^0$.}
\label{fig6}
\end{center}
\end{figure}

\begin{figure}
\begin{center}
\includegraphics[width=8cm]{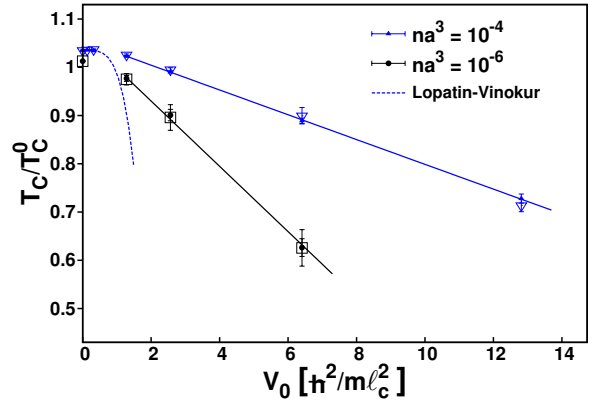}
\caption{(color online). Superfluid transition temperature as a function of the disorder strength for two values of the gas parameter $na^3$. Open and solid symbols refer respectively to $T_c$ determined from the superfluid and from the condensate fraction. The dashed line is the prediction of Ref.~\cite{Lopatin} at $na^3=10^{-4}$ shifted by $(T_c-T_c^0)$ in the absence of disorder.}
\label{fig7}
\end{center}
\end{figure}

\begin{figure}
\begin{center}
\includegraphics[width=8cm]{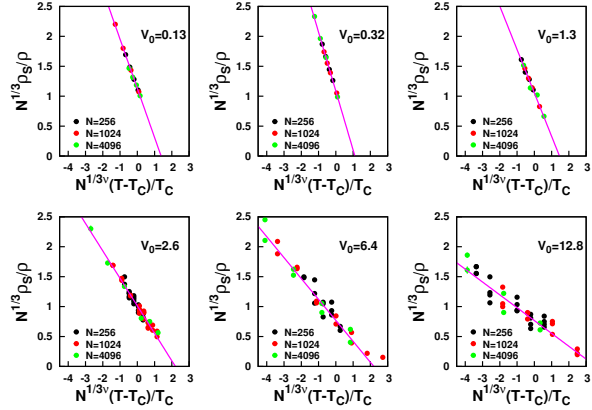}
\caption{(color online). Scaling behavior of the superfluid density for different system sizes and different realizations of disorder. The value of the gas parameter is $na^3=10^{-4}$.}
\label{fig8}
\end{center}\end{figure}

\begin{figure}
\begin{center}
\includegraphics[width=8cm]{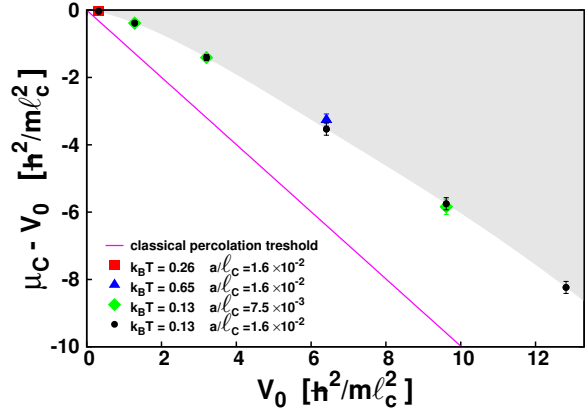}
\caption{(color online). Critical chemical potential (shifted by $V_0$) as a function of the disorder strength for different values of the temperature (in units of $\hbar^2/m\ell_c^2$) and of the scattering length. The grey shaded area denotes the superfluid phase and the (pink) solid line corresponds to the classical percolation threshold.}
\label{fig9}
\end{center}
\end{figure}

\begin{figure}
\begin{center}
\includegraphics[width=8cm]{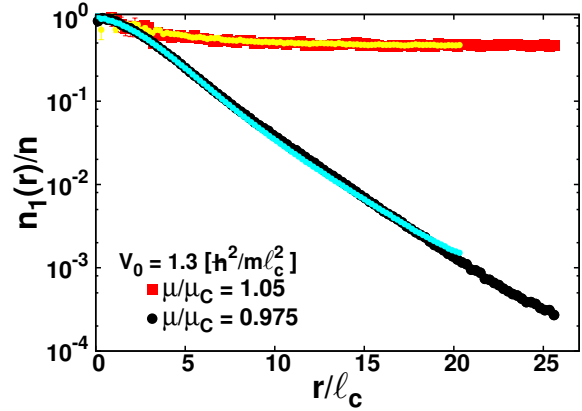}
\caption{(color online). Spatial dependence of the OBDM for two values of the chemical potential slightly below and above $\mu_c$. Here $k_BT=0.13\hbar^2/m\ell_c^2$ and $a/\ell_c=0.016$. Two different system sizes are used to check the role of finite-size effects.}
\label{fig10}
\end{center}
\end{figure}

\begin{figure}
\begin{center}
\includegraphics[width=8cm]{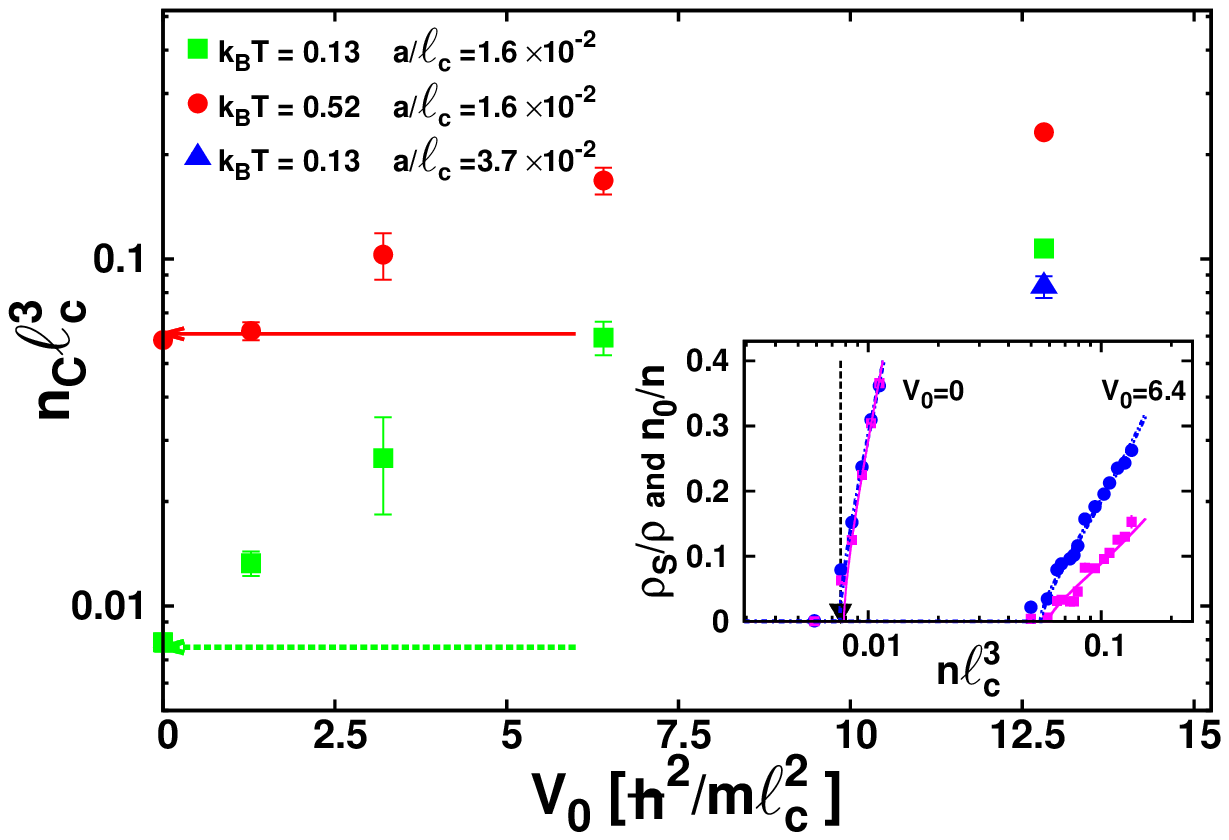}
\caption{(color online). Critical density as a function of the disorder strength for different values of the temperature (in units of $\hbar^2/m\ell_c^2$) and of the scattering length. The critical density separates the superfluid phase (above $n_c$) from the normal phase (below $n_c$). The horizontal arrows indicate the critical value $n_c^0$ of the non-interacting gas. The (blue) triangle corresponds to the same temperature but a larger scattering length compared to the (green) squares. Inset: Density dependence of $\rho_s/\rho$ [(pink) squares] and $n_0/n$ [(blue) circles] for the values $V_0=0$ and $V_0=6.4\hbar^2/m\ell_c^2$ of the disorder strength. Here $k_BT=0.13\hbar^2/m\ell_c^2$ and $a/\ell_c=0.016$. The vertical arrow indicates the corresponding value of the degenerate density $n_c^0$.}
\label{fig11}
\end{center}
\end{figure}

\section{Superfluid transition}
\label{Section4}
The effect of disorder is to suppress both the superfluid and the condensate density. In figure~\ref{fig6} we illustrate the behavior of these two quantities when the disorder strength is increased. The value of $\rho_s/\rho$ and $n_0/n$ in the absence of disorder is determined by the temperature $T$ and by the value of the gas parameter $na^3$. The figure shows a linear decrease of the superfluid and condensate components with increasing disorder in the classical regime $V_0>\hbar^2/m\ell_c^2$. An important question is to understand whether disorder also affects the superfluid transition temperature. The results are shown in figure~\ref{fig7}. The transition temperature $T_c$ is expressed in units of
\begin{equation}
T_c^0=\frac{2\pi\hbar^2}{mk_B}\left[\frac{n}{\zeta(3/2)}\right]^{2/3} \;,
\label{SUPT1}
\end{equation}
the critical temperature of the non-interacting gas with $\zeta(3/2)\simeq2.612$. In these calculations the value of the scattering length and of the disorder correlation length are kept fixed and for the latter we choose the value $\ell_c=0.6n^{-1/3}$, such that there is typically one particle in each sphere of radius $\ell_c$: $n4\pi\ell_c^3/3\simeq1$. We show results corresponding to two different densities. The reported values of $T_c$ in the absence of disorder are taken from Ref.~\cite{PGP}. At the density corresponding to the gas parameter  $na^3=10^{-4}$ we find no appreciable change of the transition temperature compared to clean systems by increasing the disorder strength up to  $V_0\sim \hbar^2/m\ell_c^2$. For larger intensities we find a sizable shift that is well described again by a linear dependence in $V_0$. For a given strength $V_0$ the reduction of the transition temperature is enhanced for smaller values of the gas parameter, consistently with the instability of the ideal Bose gas in the presence of disorder~\cite{GPS}. The value of $T_c$ is extracted from the results of the superfluid fraction $\rho_s/\rho$ ($\rho=mn$ is the total mass density), corresponding to systems with different particle number $N$, using the scaling ansatz
\begin{equation}
N^{1/3}\rho_s(t,N)/\rho=f(tN^{1/3\nu})= f(0)+f^\prime(0)tN^{1/3\nu}+... \;.
\label{SUPT2}
\end{equation}
Here, $t=(T-T_c)/T_c$ is the reduced temperature, $\nu$ is the critical exponent of the correlation length $\xi(t)\sim t^{-\nu}$, and $f(x)$ is a universal analytic function, which allows for a linear expansion around $x=0$. The validity of the scaling behavior (\ref{SUPT2}) is shown in figure~\ref{fig8}, where the effect of different realizations of the random potential is also shown. The quantity $N^{(1+\eta)/3}n_0/n$, involving the condensate fraction $n_0/n$ and the correlation function critical exponent $\eta=0.038$ of the XY-model universality class, is also expected to obey a scaling relation of the form (\ref{SUPT2}). For all reported disorder strengths $V_0$, the extracted value of the critical exponent $\nu$ is compatible with the result $\nu=0.67$ corresponding to clean systems~\cite{PGP}. It is worth noting that the values of $T_c$, obtained from the scaling law of the superfluid $\rho_s/\rho$ and of the condensate fraction $n_0/n$, coincide within our statistical uncertainty (see figure~\ref{fig7}). The presence of disorder reduces the quantum delocalization of particles occupying the deepest wells of the potential and, consequently, their contribution to the superfluid behavior. Superfluidity takes place when the degeneracy condition is met for the effectively smaller density of ``delocalized'' particles, resulting in a suppressed value of $T_c$. In Ref.~\cite{Lopatin} the shift $\delta T_c=T_c-T_c^0$ of the superfluid transition temperature is calculated using a perturbative approach for the $\delta$-correlated disorder $\langle\Delta V_{\rm{dis}}({\bf r})\Delta V_{\rm{dis}}({\bf r}^\prime)\rangle=\kappa\delta({\bf r}-{\bf r}^\prime)$, where $\Delta V_{\rm{dis}}({\bf r})=V_{\rm{dis}}({\bf r})-\langle V_{\rm{dis}}\rangle$. The $T_c$ shift is
found to be quadratic in $\kappa$, implying for our speckle potential that $\delta T_c/T_c^0=(m^2V_0^2\ell_c^3/\sqrt{na}\hbar^4)^2/[2(12\log2)^3]$, where we used a gaussian fit to the radial dependence of the autocorrelation function $\Gamma$ and considered the limit $\ell_c\to0$. We report this prediction in figure~\ref{fig7} (we also add the interaction contribution not accounted for by Ref.~\cite{Lopatin}, so that in the clean case an exact result is reproduced). Our data
in the regime of very weak disorder do not have enough precision to allow for a quantitative comparison
and diverge from the theory before $\delta T_c/T_c^0$ becomes appreciable.
The effect of disorder on the critical temperature of a hard-sphere gas was also investigated using PIMC methods in Ref.~\cite{Gordillo} where, however, no significant reduction of $T_c$ was reported. For stronger intensities of disorder, the calculation of $T_c$ becomes increasingly difficult, since the dependence on the realization gets more important (see figure~\ref{fig8}) and larger systems are needed in order to have a satisfactory self-averaging of the random potential.

In figure~\ref{fig9} we report results for the critical chemical potential $\mu_c$ obtained from calculations carried out in the grand-canonical ensemble. A small change of $\mu$ around $\mu_c$ translates into a drastic change in the long-range behavior of the OBDM (see figure~\ref{fig10}): for $\mu<\mu_c$ the OBDM decays to zero and corresponds to a normal phase, for $\mu>\mu_c$ the OBDM reaches a constant value characteristic of the superfluid state. If interactions are small but finite, we also find that the value of $\mu_c$ is essentially insensitive to a change of temperature and of interaction strength. For weak disorder, this result is accompanied by a very small critical density (see figure~\ref{fig11}) and corresponds to a renormalization of $\mu_c$ due to disorder in an extremely dilute gas. For strong disorder, it is instead consistent with the picture of a mobility edge, separating localized single-particle states from extended ones, that depends only on the parameters of the random potential.  In this latter regime we find a linear dependence of $\mu_c$ as a function of $V_0$, in agreement with the qualitative $T=0$ prediction of Refs.~\cite{Shklovskii,Nattermann} in the case of classical disorder. The figure also shows the classical percolation threshold $\mu=\epsilon_c$, whose value for the speckle potential has been determined in section~\ref{Section2}. One should notice that in the whole range of disorder intensities the critical chemical potential is significantly larger than $\epsilon_c$ as a consequence of quantum localization effects. In fact, in terms of a mobility edge, classical particles with energy larger than $\epsilon_c$ can freely move across the entire system, while in the quantum world extended states appear only for significantly larger energies bound by the inequality $\epsilon>\mu_c$.  

To conclude the study of the critical behavior, we analyze the dependence of the critical density $n_c$ on the intensity of the random potential. The calculations are carried out in the canonical ensemble at fixed temperature and scattering length. The method used to determine $n_c$ is shown in the inset of figure~\ref{fig11}. For a given value of $V_0$ one increases the density and calculates the superfluid $\rho_s/\rho$ and the condensate fraction $n_0/n$. The results are then fitted by a power-law dependence $\rho_s/\rho\sim(n-n_c)^\nu$ and $n_0/n\sim(n-n_c)^{\nu(1+\eta)}$ for $n>n_c$, where the proportionality coefficients are expected to be non-universal parameters. In the inset of figure~\ref{fig11} we show the results corresponding to a configuration without disorder ($V_0=0$) and with strong disorder ($V_0=6.4\hbar^2/m\ell_c^2$). The reported values are averaged over a few realizations of the random potential and their scatter gives an idea of the relevance of this effect. For the small value of the scattering length used here, the critical density at $V_0=0$ coincides with the non-interacting result $n_c^0=\zeta(3/2)(mk_BT/2\pi\hbar^2)^{3/2}$, while for the large $V_0$ one finds that $n_c$ is about a factor of eight greater than $n_c^0$. More comprehensive results are shown in figure~\ref{fig11} where $n_c$ is estimated from the superfluid fraction, which is less sensitive to finite-size effects. The results clearly show an increase of the critical density as a function of $V_0$, from the non-interacting degenerate density $n_c^0$ up to values $\sim 20$ times larger. It is also worth noticing that for strong disorder, an increase of the scattering length $a$ is accompanied by a decrease of $n_c$ resulting in a constant value of the critical chemical potential (see figure~\ref{fig9}).

\section{Mean-field approach}
\label{Section5}
A simple description of the thermodynamic properties of disordered systems can be provided in terms of a mean-field approach. At  $T=0$ this approach is based on the solution of the Gross-Pitaevskii (GP) equation for the order parameter in the random external field and it yields quantitatively reliable results for both the chemical potential and the density profiles. At finite temperature the mean-field theory can be efficiently applied in the case of random potentials with exceedingly long-range correlations where the local density approximation holds valid.

\begin{figure}
\begin{center}
\includegraphics[width=8cm]{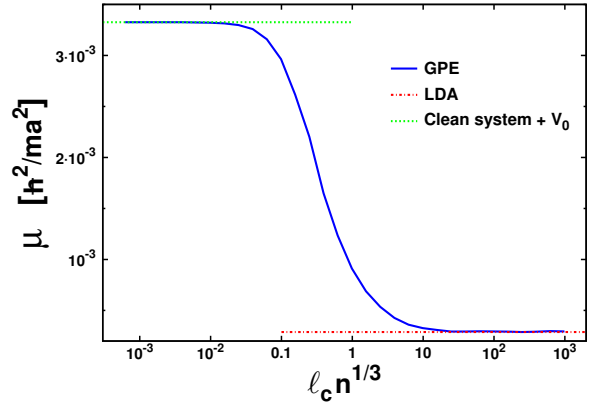}
\caption{(color online). Chemical potential at $T=0$ from the GP equation as a function of the disorder correlation length. The value of the gas parameter is $na^3=10^{-6}$ and the disorder strength is $V_0=10k_BT_c^0$. The limits of small and large $\ell_c$ are shown with horizontal lines.}
\label{fig12}
\end{center}
\end{figure}

\begin{figure}
\begin{center}
\includegraphics[width=8cm]{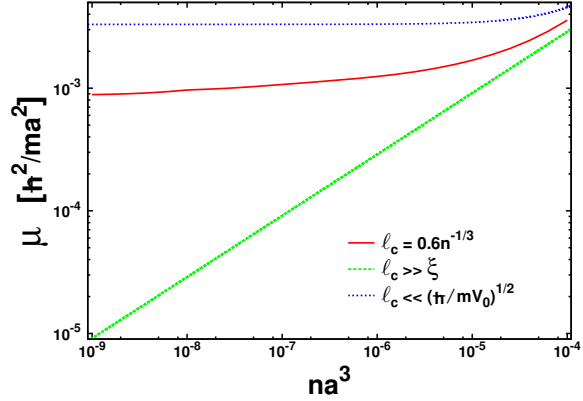}
\caption{(color online). Density dependence of the chemical potential at $T=0$ from the GP equation: (red) solid line. The (green) dashed line corresponds to the Thomas-Fermi limit $\mu=\sqrt{2gnV_0}$ and the (blue) dotted line to the opposite limit of short correlation length $\mu=gn+V_0$. The disorder strength is $V_0=3.3\times10^{-3}\hbar^2/ma^2$ and at the density $na^3=10^{-6}$ corresponds to the value used in figure\ref{fig12}.}
\label{fig13}
\end{center}
\end{figure}

\subsection{Zero temperature}
\label{Section5a}

The relevant equation is the stationary GP equation for the order parameter $\psi({\bf r})$ in the presence of the random potential
\begin{equation}
\left[ -\frac{\hbar^2}{2m}\nabla^2 + V_{\rm{dis}}({\bf r}) + g|\psi({\bf r})|^2\right] \psi({\bf r})
=\mu\psi({\bf r}) \;,
\label{GP1}
\end{equation}
where $g=4\pi\hbar^2a/m$ is the coupling constant. Within the GP approach one does not distinguish between order parameter and particle density and the following identity holds valid: $n({\bf r})=|\psi({\bf r})|^2$.  The GP equation can be obtained from the following energy functional \cite{Dalfovo}
\begin{eqnarray}
E[\psi]&=&\int d^{3}r \psi^{*}\!({\bf r})\left[ -\frac{\hbar^{2}}{2m}\nabla^2 +V_{\rm{dis}}({\bf r})\right]\psi({\bf r}) 
\nonumber\\
&+& \frac{1}{2}g|\psi({\bf r})|^4 
\label{eq:energy}
\end{eqnarray}
using the variational ansatz 
\begin{equation}
\frac{\delta}{\delta \psi^{*}} \left(E[\psi] - \mu\int d^{3}r |\psi|^2\right)=0
\label{eq:gp_ener}
\end{equation}
that corresponds to finding configurations that minimize the energy functional (\ref{eq:gp_ener}) with the normalization constraint $\int d^{3}r|\psi({\bf r})|^2=N$. The solutions of the GP equation are obtained numerically by discretizing the wave function $\psi$ on a 3D box of size $L$ (the number of grid points ranges from $64^{3}$ to $128^{3}$). Then, the energy functional $E[\psi]$ is minimized by using a conjugate gradient algorithm as described in \cite{Castin, Press}, for a given realization of the potential and a given particle density $n=N/L^{3}$. The numerical solution yields the value of the chemical potential $\mu$ as well as the spatial particle distribution.  Averages over disorder are obtained by repeating the calculation for various realizations of the random field.  

It is useful to rewrite the GP equation~(\ref{GP1}) in terms of the dimensionless variables $v_{\rm{dis}}({\bf r}) = V_{\rm{dis}}({\bf r})/V_0$, $\tilde{{\bf r}}={\bf r}/\ell_c$, $\tilde{V}_0=V_0/(\hbar^2/m\ell_c^2)$ and $\tilde{\mu}=\mu/(\hbar^2/m\ell_c^2)$. 
One finds
\begin{equation}
\left[ -\frac{1}{2}\tilde{\nabla}^2 + \tilde{V}_0v_{\rm{dis}}(\tilde{{\bf r}}) + \frac{\ell_c^2}{2\xi^2} |{\phi}(\tilde{{\bf r}})|^2\right] 
\phi(\tilde{{\bf r}})
=\tilde{\mu}\phi(\tilde{{\bf r}}) \;,
\label{GP1_1}
\end{equation}
where we have used the wavefunction $\phi({\bf r})=\psi({\bf r})/\sqrt{n}$ rescaled in terms of the average density $n$ and the healing length $\xi = 1/\sqrt{8\pi na}$. From the above equation two regimes can be investigated analytically. On the one hand, if the correlation length $\ell_c$ is much smaller than the length $\sqrt{\hbar^2/mV_0}$ fixed by the disorder strength or equivalently for weak enough disorder ($\tilde{V}_0\ll 1$), the order parameter is almost uniform $|\phi(\tilde{\bf r})|^2\simeq1$ and the effect of the random potential on the chemical potential is just a shift of the mean-field energy
\begin{equation}
\mu=gn+V_0 \;.
\label{GP2}
\end{equation}
On the other hand, if $\ell_c\gg\xi$, one can neglect the kinetic energy term in the GP equation~(\ref{GP1_1}) and use the Thomas-Fermi approximation  
\begin{equation}
\mu=gn({\bf r}) + V_{\rm{dis}}({\bf r}) \;,
\label{GP3}
\end{equation}
yielding the following result for the local particle density
\begin{equation}
n({\bf r}) = \frac{1}{g} [\mu- V_{\rm dis}({\bf r})] \Theta[\mu- V_{\rm{dis}}({\bf r})] \;.
\label{GP4}
\end{equation} 

Here $\Theta(x)$ is the Heaviside function: $\Theta(x)=1$ if $x>0$ and zero otherwise. The normalization condition $n=1/L^3\int d{\bf r} n({\bf r})$ determines the chemical potential $\mu$ in terms of the average density $n$. By using the self-averaging property (\ref{speckle5}) one obtains the equation
\begin{equation}
\frac{\mu}{V_0}+e^{-\mu/V_0}=1+\frac{gn}{V_0} \;,
\label{GP5}
\end{equation}
relating the chemical potential to the disorder strength $V_0$. 

i) If $V_0\ll gn$ then $\mu=gn+V_0$ and the disorder acts as a small shift of the pure interaction term, similarly to the short-$\ell_c$ regime of equation~(\ref{GP2}).

ii) If $V_0\gg gn$ one finds to the lowest order
\begin{equation}
\mu=\sqrt{2gnV_0} \;,
\label{GP6}
\end{equation}
corresponding to an energy per particle $E/N=2\mu/3$. 

The variation of $\mu$ with the disorder correlation length is shown in figure~\ref{fig12}, where we report the results of the GP equation for the fixed value $na^3=10^{-6}$ of the gas parameter and disorder strength $V_0=10k_BT_c^0$.  In this case $\xi n^{1/3}\sim1$ and the figure clearly shows the two limiting regimes of short and long correlation length. For the same disorder strength, in figure~\ref{fig13}, we show instead the density dependence of the chemical potential for a fixed value of $\ell_c$. The equations of state corresponding to large and small correlation lengths are also shown as a reference. 

In the regime $\ell_c\gg\xi$ one can make use of the Thomas-Fermi approximation for the order parameter which is expected to become more and more accurate as $\ell_c$ increases. Within this approximation one can make contact with the classical percolation problem of section~\ref{Section2} by noticing that, if $\mu$ is larger than the threshold energy $\epsilon_c$, the condensate density represented by equation~(\ref{GP4}) is different from zero on a percolating path ensuring thus the superfluid behavior of the system. For large $V_0$ the chemical potential increases as $\sqrt{V_0}$ according to equation~(\ref{GP6}), while $\epsilon_c$ is proportional to $V_0$. The small value of the ratio $\epsilon_c/V_0$ implies though that the quantum phase transition to the insulating state takes place only at extremely large disorder intensities.

\begin{figure}
\begin{center}
\includegraphics[width=8cm]{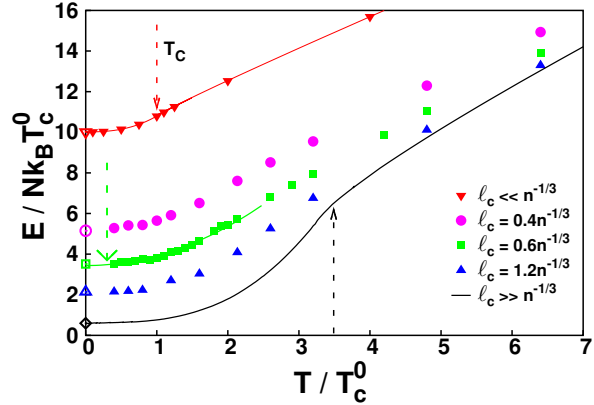}
\caption{(color online). Equation of state energy vs. temperature of a gas with interaction parameter $na^3=10^{-6}$ and disorder strength $V_0=10k_BT_c^0$. Results corresponding to different values of $\ell_c$ are reported.  The limiting cases of large correlation length: HF results with LDA (black line) and of small correlation length: clean system results shifted by $V_0$ (red symbols and line) are also shown. The open symbols at $T=0$ correspond to the results of the GP equation. The arrows indicate the superfluid transition temperature. The green line is a $T^2$ fit to the equation of state.}
\label{fig14}
\end{center}
\end{figure}

\subsection{Finite temperature}
\label{Section5b}

At $T\neq0$ one should combine the GP equation for the condensate with a proper description of the thermally excited states in the random potential. The theory becomes simple if the disorder is correlated over large distances, as one can apply the local density approximation (LDA) within standard mean-field techniques suitable for dilute gases. At low temperatures the validity condition requires $\ell_c$ to be much larger than the healing length $\xi$, while at higher temperatures the correlation length must exceed the thermal wavelength $\lambda_T$.

We use the self-consistent Hartree-Fock (HF) scheme within LDA. This mean-field approach is based on the following expression for the elementary excitations of the system in terms of their momentum ${\bf p}$ and position ${\bf r}$~\cite{JLTP}
\begin{equation}
\epsilon({\bf p},{\bf r})=\frac{p^2}{2m}+V_{\rm{dis}}({\bf r})-\mu+2gn({\bf r}) \;.
\label{FT1}
\end{equation}
The thermal density of atoms $n_T({\bf r})$ is obtained from the momentum integral of the distribution of elementary excitations
\begin{equation}
n_T({\bf r})=\int\frac{d{\bf p}}{(2\pi\hbar)^3}\frac{1}{e^{\epsilon({\bf p},{\bf r})/k_BT}-1} \;.
\label{FT2}
\end{equation}
The condensate density $n_0({\bf r})$ is determined by the finite-$T$ extension of the GP equation~(\ref{GP4}) in the Thomas-Fermi approximation
\begin{eqnarray}
 n_0({\bf r}) &=& \frac{1}{g} [\mu- V_{\rm dis}({\bf r})-2gn_T({\bf r})]  
\nonumber \\ 
&\;&\;\;\;\;\; \times\;\; \Theta[\mu- V_{\rm{dis}}({\bf r})-2gn_T({\bf r})] \;.
\label{FT3}
\end{eqnarray} 
The sum of thermal and condensate density gives the total density $n({\bf r})$, which must satisfy the overall normalization
\begin{equation}
N=\int d{\bf r} n({\bf r})=\int d{\bf r} [n_0({\bf r})+n_T({\bf r})] \;,
\label{FT4}
\end{equation}
providing the closure relation of the mean-field equations. The total energy $E$ of the system can be calculated from the following integral
\begin{eqnarray}
E&=&\int \frac{d{\bf p}\,d{\bf r}}{(2\pi\hbar)^3}\frac{p^2/2m}{e^{\epsilon({\bf p},{\bf r})/k_BT}-1}
+\int d{\bf r} V_{\rm{dis}}({\bf r})n({\bf r}) 
\nonumber\\
&+& \frac{g}{2}\int d{\bf r}[2n^2({\bf r})-n_0^2({\bf r})] \;. 
\label{FT5}
\end{eqnarray}

The HF scheme defined by the above equations neglects the quantum depletion of the condensate and at $T=0$ yields $n_0({\bf r})=n({\bf r})$, in agreement with section~\ref{Section5a}. Furthermore, it neglects the contribution of collective modes (phonons) to thermodynamics since all excitations are single particle. This approximation is known to be very accurate in dilute non-uniform systems both at high and low temperatures~\cite{JLTP}. In particular, at low temperatures, one can neglect the thermal density contribution to the expression (\ref{FT1}) of the elementary excitations and one finds the simple spectrum $\epsilon({\bf p},{\bf r})=p^2/2m+|V_{\rm{dis}}({\bf r})-\mu(T=0)|$. The term in modulus vanishes at the condensate boundaries and thermal excitations accumulate at these minima of the effective potential. The single-particle excitations close to these minima are the dominant ones at low temperature, being more important than the phonons of the bulk condensate. In fact, one can show that at low energy the single-particle excitations have the following density of states   
\begin{equation}
g(\epsilon)=\frac{4}{3}\frac{\Omega m^{3/2}}{\sqrt{2}\pi^2\hbar^3}e^{-\mu(T=0)/V_0}\frac{\epsilon^{3/2}}{V_0} \;,
\label{FT6}
\end{equation}
proportional to $\epsilon^{3/2}$ in contrast to $g(\epsilon)\propto\epsilon^2$ of phononic excitations. A similar situation occurs in harmonically trapped condensates~\cite{JLTP}.

The above semiclassical approach provides an estimate of the temperature at which Bose-Einstein condensation sets in locally in some deep well. This temperature is defined as the point where the local density $n({\bf r})$, corresponding to some deep well in the random field, reaches the critical value $n({\bf r})\lambda_T^3=\zeta(3/2)\simeq2.612$. By neglecting interactions one finds the following implicit equation for the temperature $T^\ast$
\begin{equation}
n\left(\frac{2\pi\hbar^2}{mk_BT^\ast}\right)^{3/2}=\sum_{j=1}^\infty\frac{1}{j^{3/2}}
\frac{1}{1+jV_0/k_BT^\ast} \;.
\label{FT7}
\end{equation}
The temperature $T^\ast$ is always larger than the temperature $T_c^0$ of the occurrence of Bose-Einstein in non-interacting clean systems. In particular, for large disorder strength one finds $T^\ast=T_c^0[\zeta(3/2)V_0/\zeta(5/2)k_BT_c^0]^{2/5}$. This effect comes from the reduced available volume and the corresponding higher local particle density. In the presence of weak interactions, local Bose-Einstein condensation sets in at a temperature slightly lower that $T^\ast$, because density is reduced in the wells of the random field due to repulsion and a lower temperature is needed to reach the critical value. We would like to stress that the temperature scale $T^\ast$ corresponds to the appearance of local condensates at the minima of $V_{\rm{dis}}({\bf r})$ and should not be confused with the critical temperature $T_c$ at which extended superfluidity sets in. Within the above semiclassical approach this latter temperature corresponds to the chemical potential reaching the percolation threshold of the effective potential $V_{\rm{dis}}({\bf r})+2gn_T({\bf r})$ where, according to equation~(\ref{FT3}), the condensate density $n_0({\bf r})$ is different from zero on a percolating path. For weakly-interacting systems though, because of the small value of the percolating volume fraction $\Phi_c$, the temperatures $T_c$ and $T^\ast$ are very close, unless for extremely large disorder intensities. As an example we consider the configuration shown in figure~\ref{fig14} corresponding to $na^3=10^{-6}$ and $V_0=10k_BT_c^0$ in the regime of extremely long-range correlation length $\ell_c$. The value of the temperature $T^\ast$ obtained from equation~(\ref{FT7}) is given by $T^\ast=3.63T_c^0$. The self-consistent solution of the HF equations yields a temperature $T_{\rm{BEC}}$ for the local onset of Bose-Einstein condensation in the range $3.5T_c^0<T_{\rm{BEC}}<T^\ast$. The transition temperature $T_c$ where the condensate density percolates is found to be in the range $3.4T_c^0<T_c<T_{\rm{BEC}}$. 

The HF equation of state and the corresponding transition temperature are shown in figure~\ref{fig14} for a fixed density of the gas and for large disorder strength. In the figure is also reported the equation of state corresponding to the regime of a very short correlation length $\ell_c$, where the energy is merely shifted by the average disorder intensity $V_0$ from the value of the clean system. This result is consistent with the $T=0$ prediction (\ref{GP1}) and the corresponding transition temperature coincides with $T_c$ in the absence of disorder.

\begin{figure}
\begin{center}
\includegraphics[width=8cm]{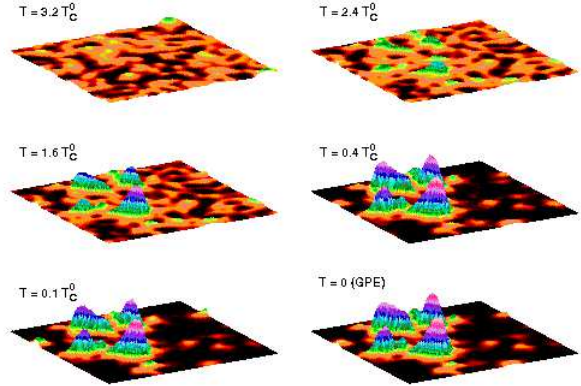}
\caption{(color online). Particle density profiles in the $(x=0,y,z)$ plane at different temperatures and for a given realization of disorder characterized by $V_0=10k_BT_c^0$ and $\ell_c=0.6n^{-1/3}$. The $T=0$ profile is obtained using the GP equation. The average density corresponds to the value  $na^3=10^{-6}$ of the gas parameter.}
\label{fig15}
\end{center}
\end{figure}

\begin{figure}
\begin{center}
\includegraphics[width=8cm]{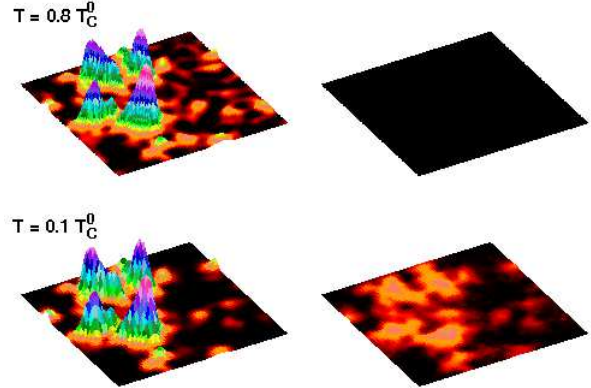}
\caption{(color online). Particle and condensate density profiles in the $(x=0,y,z)$ plane at two different temperatures, above and below $T_c$. The configuration is the same as in figure~\ref{fig15}.}
\label{fig16}
\end{center}
\end{figure}

\begin{figure}
\begin{center}
\includegraphics[width=8cm]{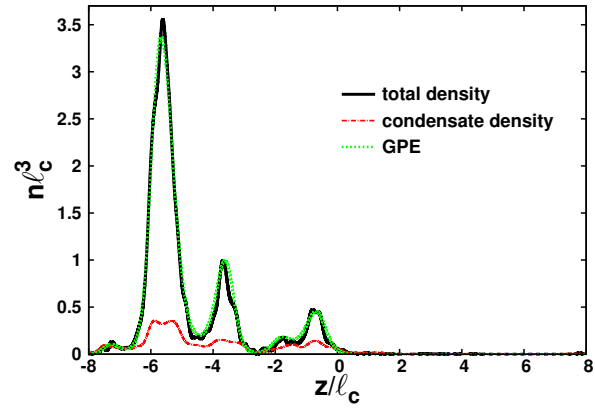}
\caption{(color online). Particle and condensate density profiles along the $(x=0,y=0,z)$ axis corresponding to the configuration of figure~\ref{fig16} at $T=0.1T_c^0$.}
\label{fig17}
\end{center}
\end{figure}

\begin{figure}
\begin{center}
\includegraphics[width=8cm]{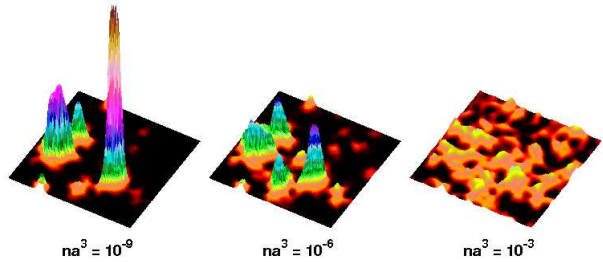}
\caption{(color online). Particle density profiles in the $(x=0,y,z)$ plane for different values of the gas parameter and for a given realization of disorder characterized by $V_0=10k_BT_c^0$ and $\ell_c=0.6n^{-1/3}$. The temperature is  $T=0.1T_c^0$.}
\label{fig18}
\end{center}
\end{figure}

\begin{figure}
\begin{center}
\includegraphics[width=8cm]{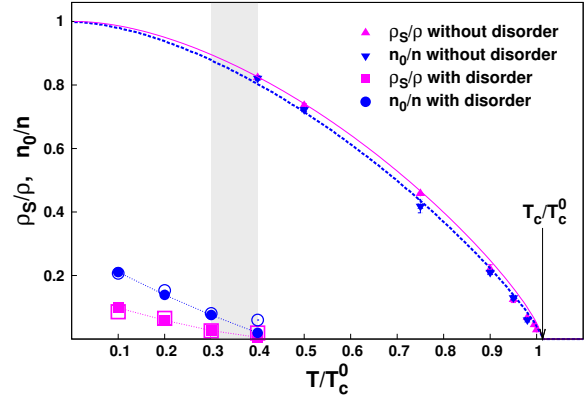}
\caption{(color online). Temperature dependence of the superfluid and condensate density without disorder and with strong disorder ($V_0=10k_BT_c^0$ and $n^{1/3}\ell_c=0.6$). The value of the gas parameter is $na^3=10^{-6}$. The grey shaded area shows the estimated range of temperatures where the superfluid transition takes place. Solid and open symbols refer to two different system sizes to check the role of finite-size effects. The solid and dashed lines in the absence of disorder are the predictions of a self-consistent mean-field calculation, while the dotted lines with disorder are guides to the eye.}
\label{fig19}
\end{center}
\end{figure}

\begin{figure}
\begin{center}
\includegraphics[width=8cm]{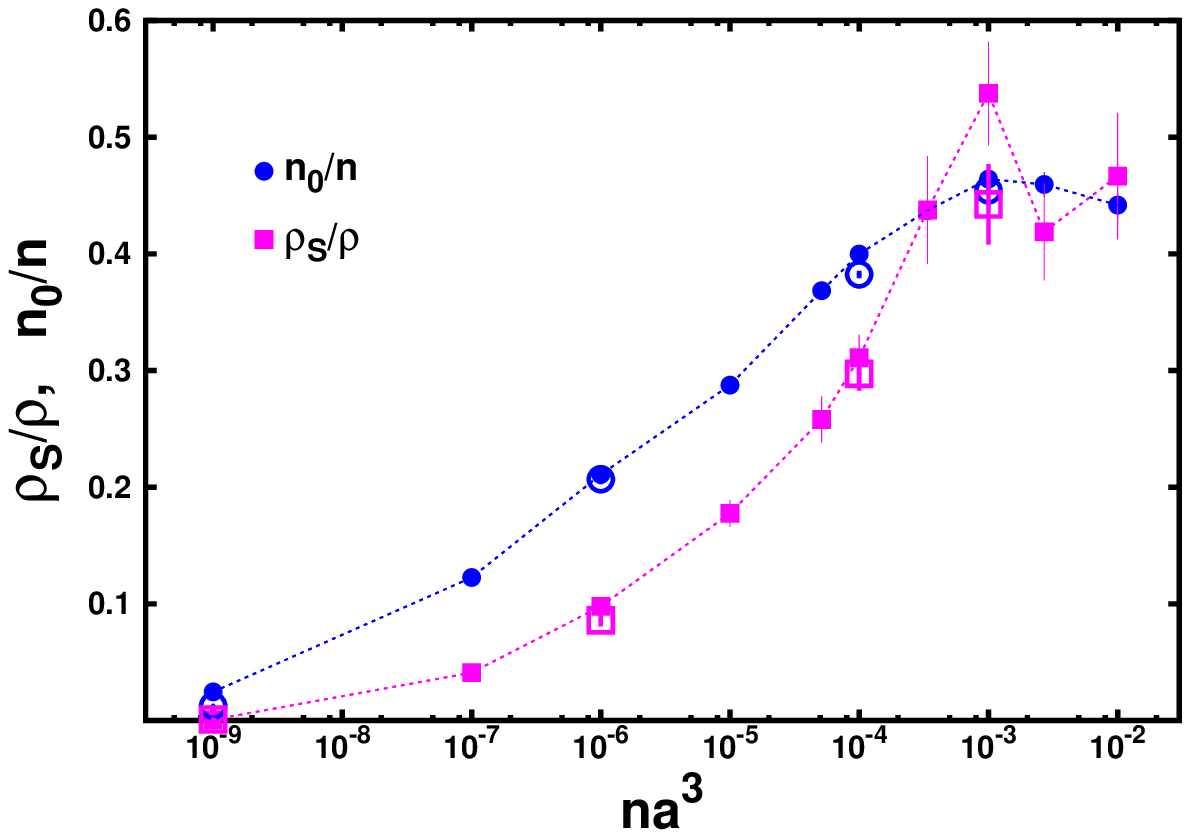}
\caption{(color online).  Dependence on the gas parameter of the superfluid and condensate density in the presence of strong disorder ($V_0=10k_BT_c^0$ and $n^{1/3}\ell_c=0.6$). The temperature is $T=0.1T_c^0$. Solid and open symbols refer to two different system sizes to check the role of finite-size effects. The error bars displayed with the solid symbols corresponding to $\rho_s/\rho$ come from averages over a number of disorder realizations. The dashed lines are guides to the eye.}
\label{fig20}
\end{center}
\end{figure}

\section{Low temperature thermodynamics}
\label{Section6}

The energy per particle obtained from PIMC simulations is reported in figure~\ref{fig14} as a function of temperature and for random potentials with different correlation lengths in the range of the mean interparticle distance. In particular, for the value $\ell_c=0.6n^{-1/3}$ we also indicate the value of the superfluid transition temperature. The equation of state and the value of $T_c$ corresponding to the limiting regimes of exceedingly long and short correlation lengths are also shown as a reference. Three important remarks about this figure are worth stressing.  a) At fixed particle density and for a given disorder strength the largest suppression of $T_c$ is achieved with $\ell_c$ on the order of the interparticle distance. b) At low temperature the energetics of the system is well described by the GP equation even when the disorder is strong and $n^{1/3}\ell_c\sim1$. c) For random potentials of large strength and with spatial correlations in the range  $n^{1/3}\ell_c\sim1$, a window of temperatures opens up where the system is normal (i.e. not superfluid), though highly degenerate ($n\lambda_T^3>1$).  

The properties of the "exotic'' normal phase displayed at low temperatures for strong disorder are worth investigating. We find that the equation of state is well fitted by a quadratic temperature dependence as  shown in figure~\ref{fig14}. This $T^2$-law for the energy, and consequently linear dependence of the specific heat, is a remarkable feature of the phase. In a clean superfluid, or in a dirty one with short-range disorder correlations (see figure~\ref{fig14}), the energy at low temperatures exhibits the $T^4$-law typical of the thermal excitation of phonons. In the opposite regime of long-range disorder correlations one can apply the HF-LDA approach described in section~\ref{Section5b}, which is expected to become more and more accurate as $\ell_c$ increases. Within this approach the system consists of large condensate lakes that may or may not be connected through a percolating path, the relevant excitations at low temperature are of single-particle nature and are localized at the boundary of the condensate lakes. These excitations contribute to the energy with a $T^{7/2}$-law\footnote{In the insulating Bose glass phase the asymptotic low-temperature law is expected to be $\sim T^2$ independent of the value of the correlation length. However, for increasing $\ell_c$, the range of temperatures where the asymptotic law applies  is suppressed.}. A Bose glass is predicted to exhibit a non-vanishing density of states at zero energy\cite{FWGF}, which results in a $T^2$-law for the low temperature equation of state.  We interpret the quadratic dependence found in the low temperature normal phase as an evidence of the Bose glass phase. 

Another important output of PIMC simulations with the random potential is the spatial particle distribution and the distribution of particles contributing to the condensate density. In figure~\ref{fig15} we show the results of the particle density $n({\bf r})$ for a given realization of the random potential. Starting from a high temperature distribution spread over the entire system, as temperature is decreased, the density becomes  more and more peaked in correspondence of the minima of the external potential. Around $T\simeq0.4T_c^0$ the system  turns superfluid (see figure~\ref{fig19}) and the density changes only slightly down to the lowest temperatures. Remarkably, the comparison with the findings of the GP equation for the same realization of the random field is rather good both for the position and the relative intensity of the peaks.  

The comparison between particle density and condensate density profiles is reported in figure~\ref{fig16} for two temperatures, above and below $T_c$. The result at the lowest temperature shows that the condensate density follows the particle distribution, but does not exhibit its pronounced peaks. This behavior is clearly represented in figure~\ref{fig17}, where we show the profiles along a cut through the plane of figure~\ref{fig16}.  Notice that the particles contributing to the condensate are only a small fraction ($\sim20\%$) of the total number of particles (see figure~\ref{fig19}). Finally, in figure~\ref{fig18} we report the density profiles as a function of the interaction strength. The figure clearly shows the effect of particle localization in the deepest wells of the random field as the value of the gas parameter is reduced.

The behavior of the superfluid and condensate fraction in the proximity of the phase transition with and without disorder are shown in figure~\ref{fig19}. It is interesting to notice the large depletion of $\rho_s$ and $n_0$ even at the lowest temperatures and the fact that, in the regime of strong disorder, the superfluid component becomes significantly smaller than the condensate one. Similar results are reported in figure~\ref{fig20} for a fixed temperature and for varying values of the interaction strength. The figure clearly shows that by making interactions weaker the system will eventually turn normal and that the difference between superfluid and condensate fraction disappears  with strong enough interactions.  

To complete the picture we would like to mention the limit of weak disorder $(V_0 m\ell_c^2/\hbar^2) \ll 1$ and weak interactions $na^3 \to 0$ considered in Ref.~\cite{Nattermann}. It explains how interactions stabilize the superfluid phase starting from the localized phase of the non-interacting gas. At zero temperature the boundary between the insulating (Bose glass) and superfluid phases is set (up to logarithmic corrections) by percolation at the Larkin energy scale $ V_0^4 m^3 \ell_c^6/ \hbar^6 $ which has to be compared with the chemical potential of the weakly interacting gas $gn$. This leads to the relation for the critical density $(na)_c \sim V_0^4$, {\it i.e.} for small $V_0$ only a tiny density of particles is required to off-set the localization effects. In practice, one observes extremely robust superfluidity of the weakly-interacting Bose gas even in the presence of relatively large disorder, see Figs.~\ref{fig7}, \ref{fig19}, \ref{fig20}; for $na \gg (na)_c$ the critical temperature is essentially the same as in the ideal Bose gas.

\section{Conclusions}
\label{Conclusions}
We find that in a quantum degenerate bosonic gas a random potential is most efficient in suppressing superfluidity if it is correlated over length scales comparable with the mean interparticle distance. However, for the typical diluteness conditions of ultracold gases, disorder intensities significantly larger than the energy scale $k_BT_c^0$ set by the BEC transition of the ideal gas are required for a significant reduction of the superfluid critical temperature and stronger interactions make the superfluid state more robust. In the regime of weak interactions and strong disorder, the superfluid transition turns out to be  well characterized by the existence of a mobility edge, separating localized from extended states, that is largely independent of temperature and interaction strength. This picture is similar to the percolation threshold of classical particles, but we find that quantum localization effects drive the system normal in a large region of energies where classical states would be extended. Furthermore, most of the particles are localized in the deepest wells of the random potential and only a small proportion contributes to the extended condensate state. The effective density of these delocalized particles is much smoother due to the screening of the external field from the other particles and sets the critical density of superfluidity which however starts at significantly lower temperatures compared to clean systems.  For larger disorder intensities an "exotic'' normal phase appears in the degenerate regime, even though we can not reach $T=0$. This phase is characterized by a peculiar $T^2$ dependence of the equation of state, that is markedly different from the $T^4$ law of homogeneous superfluids and from the $T^{7/2}$ law found for large condensate lakes within LDA and is in agreement with the predictions for the Bose glass phase. Remarkably, some aspects of this phase, such as the $T=0$ equation of state and the spatial distribution of particles can be correctly described using the mean-field GP theory.

\section*{Acknowledgements}
We acknowledge useful discussions with B. Svistunov, and L.P. Pitaevskii. This work, as part of the European Science Foundation EUROCORES Program ``EuroQUAM-FerMix'', was supported by funds from the CNR and the EC Sixth Framework Programme. NP acknowledges support from NSF grant PHY-0653183. Calculations have been performed on the HPC facility {\it Wiglaf} at the Physics Department of the University of Trento and on the BEN cluster at ECT$^{\ast}$ in Trento.

\end{document}